\documentclass{bulsao}
\usepackage{amssymb}
\usepackage{amsmath}
\usepackage{graphicx}
\usepackage{latexsym}
\usepackage{longtable}

\begin{document}

\title{Speckle Interferometry of Metal-Poor Stars in the Solar Neighborhood.\,II}

\author{D.~A.~Rastegaev, Yu.~Yu.~Balega, A.~F.~Maksimov, E.~V.~Malogolovets, V.~V.~Dyachenko}

\institute{Special Astrophysical Observatory, RAS, Nizhnii Arkhyz,
Karachai-Cherkessian Republic, 357147 Russia}

\offprints{D.~A.~Rastegaev, \email{leda@sao.ru}}

\date{received: April 7, 2008/revised: April 21, 2008}

\titlerunning{Speckle Interferometry of Metal-Poor Stars .II}
\authorrunning{Rastegaev et al.}

\abstract{
The results of speckle interferometric observations of 115
metal-poor stars $(\mathrm{[m/H]}<-1)$  within 250~pc from the Sun
and with proper motions $\mu \gtrsim 0.2''/$yr, made with the 6-m
telescope of the Special Astrophysical Observatory of the Russian
Academy of Sciences, are reported. Close companions with
separations ranging from  $0.034''$ to  $1''$ were observed for 12
objects --- G76-21, G59-1, G63-46, \mbox{G135-16,} G168-42, G141-47, G142-44, G190-10, G28-43, G217-8, \mbox{G130-7}, and G89-14 --- eight of them are astrometrically resolved for the first time. The newly resolved systems include one triple star --- G190-10. If combined with spectroscopic and visual data, our results imply a single:binary:triple:quadruple star ratio of 147:64:9:1 for a sample of 221 primary components of halo and thick-disk stars.
}

\maketitle

\section{INTRODUCTION}

Metal-poor stars of the Galactic halo and thick disk bear important information about the chemical and kinematical properties of matter at the epoch of the formation of the Milky Way. Of special importance is the study of the orbital parameters of binary and multiple systems, which provide a source of data
on stellar masses and luminosities.

\begin{figure}[tbp]
\includegraphics[width=8.5cm]{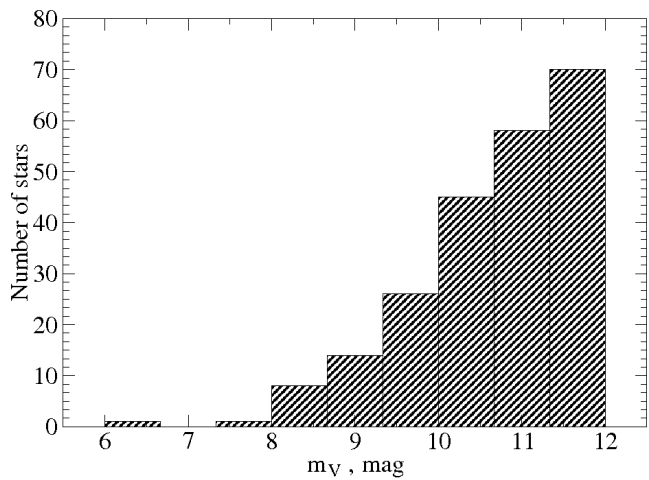}
\caption{Distribution of the $V$-band magnitudes of stars of the sample studied.}
\label{V_mag:Rastegaev_n}
\end{figure}

\begin{figure}[tbp]
\includegraphics[width=8.5cm]{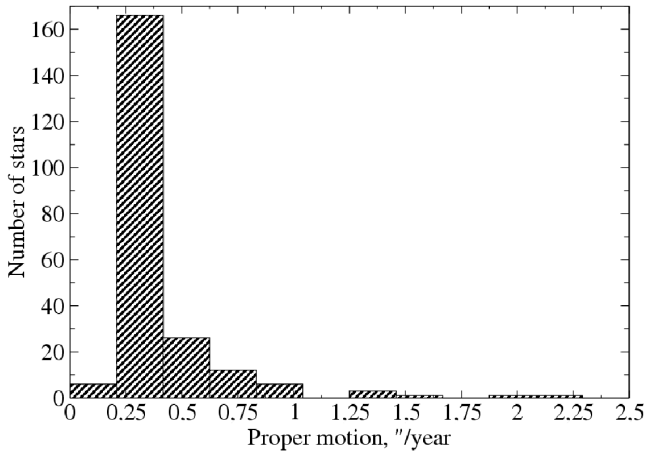}
\caption{Distribution of the proper motions
of the stars of the sample. For better visualization, we do not
show in the histogram the star G122-51 with anomalously high
proper motion ($\mu=7.042\ ''/$yr).} \label{pm:Rastegaev_n}
\end{figure}

To estimate the fraction of multiple stars and determine the
orbital parameters of old metal-poor stars, we started a
speckle interferometric survey of such objects located within 250~pc from the Sun. Rastegaev et al. (2007)
described a sample of 223 population-II dwarf stars in the solar
neighborhood and reported the results of the survey of the first 109 stars of the sample performed with the 6 m telescope of the Special Astrophysical Observatory of the Russian Academy of Sciences (SAO RAS). The sample includes nearby F, G, and early K-type subdwarfs down to 12th magnitude in the  $V$ band \mbox{(Fig. \ref{V_mag:Rastegaev_n})} with metalicities  $\mathrm{[m/H]}<-1$ and proper motions $\mu \gtrsim 0.2''/$yr \mbox{(Fig. \ref{pm:Rastegaev_n}).} In this paper we continue to report the results of speckle interferometric observations for the remaining 114 stars of the halo and thick disk performed with the 6 m telescope of the SAO RAS in 2007. We also report the results of repeated speckle-interferometric observations of our earlier resolved subsystem of the quadruple star \mbox{G89-14} (\cite{rastegaev:Rastegaev_n}).

\vspace*{0.3cm}
\section{OBSERVATIONS}

The speckle-interferometric observations were performed with the
6 m telescope of the  of the SAO RAS
 in March (52 objects), June--July (52
objects), and September (10 objects), 2007. In addition, we also
reobserved in two filters the interferometric subsystem of the quadruple star
\mbox{G89-14} that we discovered in December, 2006
(\cite{rastegaev:Rastegaev_n}). In September we observed six
objects resolved in June and July (G141-47,
G142-44, G217-8, G130-7,G190-10, and G28-43). We also reobserved
the unresolved objects G183-9,  G24-17, G26-1, and G128-11 in
order to obtain their power spectra with higher signal-to-noise
ratio.

In our observations we use a facility based on  EMCCD (a
\mbox{512$\times$512} CCD with internal electron gain), which has
high quantum efficiency and linearity, allowing objects to be
discovered with component magnitude differences $\triangle m
\lesssim 5$ at the diffraction-limited resolution of the
telescope. The size of the detector field (4.4$''$) allowed
secondary components to be discovered at separations as large as
3$''$ from the primary star.

We recorded the speckle interferograms in three filters: $550/20$, $600/40$, and $800/100$ nm
(the numbers indicate the central transmission wavelength of the filter and the transmission bandwidth,
respectively) with exposures ranging from 5 to 20 milliseconds. We took 2000 images in each filter
for almost all the objects observed in March and 1940 exposures for every object observed in June, July,
and September.

Weather conditions during March set were not favorable for speckle-interferometric observations (seeing
was at about  $3''$). During the June and September observations, on the contrary, seeing was \mbox{$(1.0-1.5)''$,}
and sometimes even better than $1''$.

We calibrated our measurements using the so-called  ``standard''
pairs---binaries with well-known component separations and
position angles. In addition, in September we used an opaque mask
with a pair of circular holes, which was located in the beam
converging from the primary mirror of the telescope, to
calibrate the scale and position angle. The known geometry of the
holes allows the image scale and angular orientation of the CCD to
be determined in each filter from the fringe pattern. In this
method we used Deneb as the bright source.

Descriptions of the technique that we used to determine the
relative positions and component magnitude differences of the
objects studied from speckle interferograms averaged over a series
of power spectra can be found in Balega et al. (2002) and Pluzhnik (2005). The accuracy of
this technique may be as good as  $0.02^m$, $0.001''$, and $0.1
\degr$ for the component magnitude difference, separation, and
position angle, respectively.

\section{RESULTS OF OBSERVATIONS}
Table~ \ref{tab1:Rastegaev_n} lists the resolved stars.
We observed the speckle-interferometric components for
12 objects, \mbox{one of them} is the quadruple system G89-14,
which was resolved for the first time in December, 2006
(\cite{rastegaev:Rastegaev_n}). We were the first to
astrometrically resolve eight  (G76-21, G135-16, G141-47, G142-44,
G190-10, G28-43, G217-8, and G130-7) of the 11 remaining systems.
Among the new systems we point out  G190-10, where we discovered
the third component in the earlier known spectroscopic \mbox{binary
(\cite{paperXVI:Rastegaev_n}).}

\bigskip

In addition to new observations, we also report the results of our
re-reduction of stars from (\cite{rastegaev:Rastegaev_n}) \mbox{(Table~
\ref{tab2:Rastegaev_n}).} Our new results for these stars slightly differ
 from those reported in our previous
paper (\cite{rastegaev:Rastegaev_n}) by corrected estimates of
angular separations due to the refined coefficients of the
transition from pixel measurements to angular units. It goes
without saying that these corrections have appreciable effect only
for wide pairs. In addition, we also thoroughly analyzed the
measurement errors for each object and, in contrast to our earlier
paper (\cite{rastegaev:Rastegaev_n}), give the epoch of observation
for each individual pair. We list the so far unresolved stars in
Table~ \ref{unresolved:Rastegaev_n}.

\section{SUPPLEMENTARY DATA FOR RESOLVED STARS}

In this section we gather the supplementary data on resolved stars
(see also Table~\ref{additional:Rastegaev_n}). For some of the
objects we give two distances inferred from  \mbox{trigonometric
(\cite{esa:Rastegaev_n})} and photometric (\cite{clla:Rastegaev_n})
parallaxes. The latter distance is evidently underestimated,
because it does not take the additional component into account. On
the other hand, the additional component also contributes to the
error of the measured trigonometric parallax, especially in
short-period systems.

{\bf G76-21} ($02^h41^m13\fs6$ $+09\degr 46' 12''$;  HIP~12529).
It is an F2-type star (\cite{simbad:Rastegaev_n}) at a heliocentric distance of about 190~pc (\cite{esa:Rastegaev_n}) or 90~pc according \mbox{to Carney et al. (1994).} The star was observed using the method of lunar occultations, but it was not resolved into individual components
(\cite{richichi2002:Rastegaev_n}). It is known as a suspected SB2
system based on the results of metallicity measurements
(\cite{clla:Rastegaev_n}). Spectroscopic observations of this star
show signs of about 10-day periodicity, however, no conclusive
evidence could be found for radial-velocity variations
(\cite{latham_pc:Rastegaev_n}). We were the first to
astrometrically resolve this star.

{\bf G89-14} ($07^h22^m31\fs5$ $+08\degr 49' 13''$; HIP~35756;
WDS~07224+0854). We earlier discovered the fourth component
(\cite{rastegaev:Rastegaev_n}) in  the triple system consisting of
a spectroscopic binary with a period of 190 days
(\cite{paperXVI:Rastegaev_n}) and a common proper-motion companion
at an angular  separation of $34''$ from this binary
(\cite{allen:Rastegaev_n}). Repeated observations in the 550/20
filter failed to reveal the speckle interferometric component at a
distance of $0.98''$ from the spectroscopic pair, because in this
part of the spectrum the component in  question is 5$^m$ fainter
than the SB1 pair. Observations in the 800/100 filter (see Table
\ref{tab1:Rastegaev_n}) performed in March, 2007, confirmed the
results of the December, 2006 observations to within the quoted
errors. The heliocentric distance of the quadruple system is equal
to \mbox{$95$~pc (\cite{clla:Rastegaev_n})} or to about 170~pc
according to (\cite{esa:Rastegaev_n}).

{\bf G59-1} ($12^h08^m54\fs7$ $+21\degr 47' 19''$; HIP~59233;
WDS~12089+2147). It is a triple system. The inner pair, which has
an integrated spectral \mbox{type of
G2V(\cite{simbad:Rastegaev_n})} was discovered by \emph{HIPPARCOS}
(\cite{esa:Rastegaev_n}). The outer component, which is located at
an angular distance of about $16''$, has common proper motion with
the inner pair (\cite{allen:Rastegaev_n}). We resolved the inner
subsystem. The heliocentric distance is equal to about 110~pc
(\cite{esa:Rastegaev_n}) or 50~pc according to
(\cite{clla:Rastegaev_n}).

{\bf G63-46} ~~($13^h39^m59\fs6$ $+12\degr 35' 22''$;~~ HIP~66665;
WDS~13400+1235). A double star of spectral type F9V
(\cite{simbad:Rastegaev_n}) was first resolved by the~
\emph{HIPPARCOS} satellite (\cite{esa:Rastegaev_n}) and
measured speckle interferometrically by Zinnecker et al.
(2004) and Hartkopf et al. (in preparation). The
heliocentric distance of the system is equal to about 130~pc
(\cite{esa:Rastegaev_n}) or 60~pc (\cite{clla:Rastegaev_n}).

{\bf G135-16} ($14^h04^m01\fs6$ $+22\degr 31' 30''$; HIP~68714). A
double star of spectral type G2 (\cite{simbad:Rastegaev_n}). This
pair, which we astrometrically resolved for the first time, must
be an SB1 type spectral binary with a period of 2606 \mbox{days
(\cite{paperXVI:Rastegaev_n}).} Its estimated heliocentric distance is
about  80~pc (\cite{esa:Rastegaev_n}) or \mbox{65~pc
(\cite{clla:Rastegaev_n}).}

{\bf G168-42} ($16^h19^m51\fs7$ $+22\degr 38' 20''$; HIP~80003).
An astrometric binary (\cite{zinnecker:Rastegaev_n,law:Rastegaev_n}) of spectral type \mbox{sd:G2 (\cite{simbad:Rastegaev_n}).} We are the first to
report the component magnitude difference for this system. Latham et al. (2002) list it as a spectroscopic binary with unknown period. The star exhibits a systematic decrease of radial velocity over more than 24 years of its spectroscopic observations(\cite{latham_pc:Rastegaev_n}). Its heliocentric distance is about 110~pc (\cite{esa:Rastegaev_n}) or
100~pc (\cite{clla:Rastegaev_n}) if inferred from the trigonometric
or photometric parallax, respectively.

{\bf G141-47} ($18^h53^m16\fs5$ $+10\degr 37' 26''$; BD+10\degr
3711
 TYC~1030-316-1).
This first resolved pair with an angular separation of about $0.04''$
is an SB1
spectroscopic binary with a period of 388.52 \mbox{ days
(\cite{paperXVI:Rastegaev_n})} and a spectral type of F8
(\cite{simbad:Rastegaev_n}). We may have discovered the third
component in the known spectroscopic pair. The heliocentric
distance to this object is 110~pc (\cite{clla:Rastegaev_n}). The
system was observed twice in June in the  550/20 and 800/100
filters, and also in September in the 800/100 filter.
In Table~ \ref{tab1:Rastegaev_n} we give only the preliminary photometry of the speckle interferometric pair
based on the results of the June observations in the 800/100 filter due to the low signal-to-noise ratio
of the integrated power spectra.

{\bf G142-44} ($19^h38^m53\fs2$ $+16\degr 25' 34''$; NLTT~48059;
TYC~1602-2423-1). This first resolved G5-type binary
is located at a heliocentric distance of 110~pc
(\cite{clla:Rastegaev_n}). We observed this pair four times (see
Table \ref{tab1:Rastegaev_n}), and three of them in the 800/100
filter. The weak fringe contrast in the power spectrum in the
600/40 filter allowed only the lower boundary of component
magnitude difference to be estimated in this part of the spectrum.

{\bf G190-10} ($23^h07^m59\fs8$ $+41\degr 51' 20''$; NLTT~55914;
TYC 3224-2564-1). A new triple system of spectral type G1
(\cite{simbad:Rastegaev_n}). We found the third, outer component at
an angular distance of $0.98''$ from this earlier known SB1
system with a period of 30 days (\cite{paperXVI:Rastegaev_n}). The
object is located at a distance of 90~pc
(\cite{clla:Rastegaev_n}).

{\bf G28-43} ($23^h09^m32\fs9$ $+00\degr 42' 40''$;  HIP~114349).
A binary of the spectral type G2 (\cite{simbad:Rastegaev_n}), which
we resolved for the first time. The object is located at a
distance of  40~pc (\cite{clla:Rastegaev_n}). The
\emph{HIPPARCOS} catalog lists no parallax for the system
(\cite{simbad:Rastegaev_n}). The wide component CCDM J23096+0043B
at an angular separation of $12.2''$ does not form a physical
pair (\cite{osorio:Rastegaev_n}).

{\bf G217-8} ($23^h26^m32\fs8$ $+60\degr 37' 43''$;  HIP~115704).
We were the first to astrometrically resolve this F2-type
spectroscopic binary  (\cite{simbad:Rastegaev_n}) with a
preliminary orbit (9632 days period)
(\cite{paperXVI:Rastegaev_n}). The distance to the system is equal
to about \mbox{110~pc (\cite{esa:Rastegaev_n})} or \mbox{105~pc
(\cite{clla:Rastegaev_n}).} We observed the object twice: in June,
in the 600/40 filter and in September, in the 800/100 filter.
Unfortunately, the insufficient quality of the power spectra in
both filters did not make it possible to determine the component
magnitude difference and showed up in the accuracy of the inferred
positional parameters \mbox{(see Table \ref{tab1:Rastegaev_n}).}

{\bf G130-7} ($23^h45^m00\fs1$ $+30\degr 20' 10''$;  HIP~117150).
An F-type system (\cite{simbad:Rastegaev_n}) at a
distance of about 160~pc (\cite{esa:Rastegaev_n}) (or 120~pc from other data
(\cite{clla:Rastegaev_n})), which was resolved
for the first time.

\section{MULTIPLICITY OF STARS}
\subsection{Distant Components}

We use the additional available data on spectroscopic multiplicity
(\cite{paperXV:Rastegaev_n,paperXVI:Rastegaev_n}) and distant
components from the  \mbox{WDS (\cite{wds:Rastegaev_n})} for the
114 objects studied. Whereas spectroscopic and interferometric
measurements provide conclusive evidence indicating that the
components in question are physically bound, wide visual
components should be treated with more care. We found a total of
104 WDS companions for our stars and discarded most of them as
optical projections. Table \ref{wds:Rastegaev_n} lists the data
for all the wide components found for the stars of our sample.
Column~1 gives the names of the stars studied; column~2---all the
WDS components found for the star. For the components found to be
physically bound to the stars of the sample columns~3 and 4 list
the angular separation (in arcsec) and magnitude difference,
respectively. Column~5, which is entitled ``Status'', indicates
the components that we consider to be physically bound with the
primary star (``$+$'') and optically projected (unbound) pairs
(\mbox{``$-$''}). The additional question mark in this column
indicates that we are not certain about the adopted decision,  nd
a single question mark indicates that only a single measurement is
available, which does not allow any conclusions concerning the
physical bound between the components. The last column gives the
references to the papers containing the data on the corresponding
pair and whether the physical association between the components
is confirmed or disproved. These are in all cases the papers of
Allen et al. (2000) and \mbox{Zapatero~Osorio \& Martin (2004)} dedicated to wide pairs of
population-II stars and the HIPPARCOS catalog (\cite{esa:Rastegaev_n}). The additional $\star$ symbol in this
column indicates that our observations confirm the presence of the
component considered. In  cases with no references  given
we made  decision concerning the physical boundness on our own,
based on the data listed in the WDS catalog. To this end, we
analyzed the variations of component separations and magnitude
differences with time.

As a result, we left only seven WDS components (marked ``$+$'' or ``$+?$'') of  104, and took them
into account when counting the number of systems of different multiplicity.

\subsection{Ratio of systems of different multiplicity}
We computed the ratio of systems of different multiplicity using
all the published data on the observations of the corresponding
systems using various methods. Of the 114 stars considered 27 are
spectroscopic binaries
(\cite{paperXV:Rastegaev_n}; \cite{paperXVI:Rastegaev_n}; \cite{clla:Rastegaev_n}) and 11 stars are speckle interferometric binaries. Seven stars have companions from the WDS catalog. When analyzing spectroscopic binaries we took into account both the pairs with known orbital periods and the systems for which no  periods have been determined. It goes without saying that there exist components which can be found
using several different methods. In addition, we also analyzed the
ratio of systems of different multiplicity from Rastegaev et al.
(2007). We excluded  \mbox{G120-15} from the list of binary stars, because only one measurement of the positional parameters is available for this star, which does not allow any conclusions
be made for it. We added  two unaccounted binaries
BD+25$\degr$~1981 and HD97916 from (\cite{paperXIV:Rastegaev_n}),
and assumed the three systems---\mbox{G43-3} (see also
(\cite{paperXIV:Rastegaev_n})), \mbox{G186-26,} and G210-33 to be binaries
based on the small variation of radial velocities
(\cite{latham_pc:Rastegaev_n}).

As a result, the single:binary:triple:quadruple ratio for the 221
primary-component halo and thick-disk stars
(\cite{rastegaev:Rastegaev_n}) discovered using all methods is
equal to  147:64:9:1. Thus out of  306 stars considered---223
observed stars and 83 their satellites---more than a half (159)
belong to multiple systems. The multiplicity of the sample---i.e.,
the ratio of the number of multiple systems to the total number of
systems---is about 33\%.

Duquennoy \& Mayor (1991) obtained a similar
estimate for disk stars of spectral types ranging from F7 to G9
and found it to be 51:40:7:2. We point out the difference between
the two samples compared. Whereas we constructed our sample by
selecting stars with certain magnitudes and space velocities, the
sample of Duquennoy and Mayor is only
distance limited: all their stars are located within 22~pc from the
Sun.

\section{CONCLUSIONS}
The speckle interferometric survey of 223 metal-poor stars from the solar
neighborhood was performed with the 6 m telescope of the Special Astrophysical
Observatory. Nineteen binary and multiple systems were resolved. From these,
15 objects were resolved astrometrically for the first time.
Three of our resolved systems---G76-21 (HIP~12529), \mbox{G114-25}
(HIP~44111), and G217-8 (HIP~115704) have metallicities
$\mathrm{[m/H]}<-2$ (\cite{clla:Rastegaev_n}). The additional data
on the spectroscopic
(\cite{paperXV:Rastegaev_n}; \cite{paperXVI:Rastegaev_n}; \cite{clla:Rastegaev_n}; \cite{paperXIV:Rastegaev_n})
and astrometric
(\cite{wds:Rastegaev_n}; \cite{osorio:Rastegaev_n}; \cite{allen:Rastegaev_n})
multiplicity allowed us to estimate the
single:binary:triple:quadruple star ratio to be 147:64:9:1.

Part of the speckle interferometric pairs with relatively short periods are suitable for monitoring in order
to compute their orbits and determine the masses of metal-poor stars, which are necessary for the calibration of the
mass--luminosity relation. Such studies are of great importance, because even now we badly lack the empirical data
for the metallicity interval considered.

The sample presented in (\cite{rastegaev:Rastegaev_n}) is the most
thoroughly analyzed one in terms of the multiplicity of halo and
thick-disk stars. This circumstance allows the sample to be used
for statistical studies where physically bound components play
important part. One must bear in mind
the selection effects due to heliocentric distances to the
objects, their multiplicity and proper motions.
An addition, low-mass companions could be missed for the stars of survey
because of limitations of the methods.
All this must stimulate further observations and theoretical studies.

\begin{acknowledgements}
This work was supported by the Russian Foundation for Basic
Research (project no.~04-02-17563) and the program
of the  Physical Sciences of the Russian Academy of Sciences. This
research made use of the Simbad
database and  \mbox{WDS (\cite{wds:Rastegaev_n})} catalog. We are grateful
to D.Latham for sharing the data on the spectroscopic multiplicity
and orbital periods of selected \mbox{objects.}
\end{acknowledgements}

\begin{table*}
\begin{center}
\caption{Results of the speckle-interferometric measurements for resolved objects}\label{tab1:Rastegaev_n}
\begin{tabular}{l | l | c | c | c | c | c | c | c | c}
\hline
\textbf{Name} & \textbf{Other} & \textbf{Epoch} &$\boldsymbol{\Theta (\degr )}$ & $\sigma_{\Theta}$ & $\boldsymbol{\rho\ ('')}$ & $\sigma_{\rho}$ & $\boldsymbol{\bigtriangleup m}$ & $\sigma_{\boldsymbol{\bigtriangleup m}}$ & $\lambda / \bigtriangleup \lambda$ \\
\textbf{of the object}  & \textbf{designation} &     &     &     &     &     &     &     &  \\
\hline
G76-21    & HIP~12529         & 2007.73116  & 206.4 & 0.7 & 0.047 & 0.001 & 0.4  & 0.1  & 800/100   \\
G89-14    & HIP~35756         & 2007.24040  & 0.8   & 0.4 & 0.982 & 0.005 & 4.3  & 0.1  & 800/100   \\
G59-1     & HIP~59233         & 2007.24333  & 280.5 & 0.7 & 0.098 & 0.002 & 1.4  & 0.1  & 800/100   \\
G63-46    & HIP~66665         & 2007.24084  & 82.9  & 0.3 & 0.222 & 0.001 & 0.94 & 0.02 & 550/20    \\
G135-16   & HIP~68714         & 2007.24388  & 174.8 & 1.7 & 0.034 & 0.001 & 0.7  & 0.1  & 550/20    \\
G168-42   & HIP~80003         & 2007.24109  & 208.0 & 0.4 & 0.180 & 0.001 & 1.34 & 0.02 & 800/100   \\
G141-47   & BD+10 \degr ~3711 & 2007.48727  & 143   & 4   & 0.041 & 0.005 & 0.9  & 0.6  & 800/100   \\
G141-47   & BD+10 \degr ~3711 & 2007.48728  & 139   & 20  & 0.034 & 0.013 &      &      & 550/20    \\
G141-47   & BD+10 \degr ~3711 & 2007.73870  & 137   & 6   & 0.044 & 0.005 &      &      & 800/100   \\
G142-44   & NLTT~48059        & 2007.49008  & 193.2 & 0.7 & 0.661 & 0.007 & 3.7  & 0.2  & 800/100   \\
G142-44   & NLTT~48059        & 2007.49286  & 192.9 & 0.5 & 0.663 & 0.005 & 3.7  & 0.1  & 800/100   \\
G142-44   & NLTT~48059        & 2007.49287  &       &     &       &       &      &      & 600/40    \\
G142-44   & NLTT~48059        & 2007.73871  & 193.3 & 0.5 & 0.665 & 0.005 & 3.85 & 0.06 & 800/100   \\
G190-10   & NLTT~55914        & 2007.51184  & 287.0 & 0.2 & 0.977 & 0.001 & 1.39 & 0.02 & 800/100   \\
G190-10   & NLTT~55914        & 2007.73885  & 286.9 & 0.3 & 0.982 & 0.002 & 1.37 & 0.02 & 800/100   \\
G190-10   & NLTT~55914        & 2007.73886  & 286.9 & 0.3 & 0.982 & 0.002 & 1.73 & 0.03 & 550/20    \\
G28-43    & HIP~114349        & 2007.51209  & 37.6  & 0.4 & 0.425 & 0.003 & 3.35 & 0.04 & 800/100   \\
G28-43    & HIP~114349        & 2007.51210  & 37.4  & 0.6 & 0.425 & 0.004 & 3.5  & 0.1  & 600/40    \\
G28-43    & HIP~114349        & 2007.73877  & 37.7  & 0.4 & 0.424 & 0.003 & 3.32 & 0.03 & 800/100   \\
G217-8    & HIP~115704        & 2007.49527  & 263   & 5 & & 0.09  & 0.02         &      & 600/40    \\
G217-8    & HIP~115704        & 2007.72510  & 260   & 5 & & 0.07  & 0.02         &      & 800/100   \\
G130-7    & HIP~117150        & 2007.51188  & 230.0 & 1.5 & 0.191 & 0.005 & 2.98 & 0.06 & 800/100   \\
G130-7    & HIP~117150        & 2007.73888  & 230.7 & 1.0 & 0.191 & 0.004 & 2.95 & 0.04 & 800/100   \\
\hline
\end{tabular}
\end{center}
\end{table*}

\begin{table*}
\begin{center}
\caption{Speckle-intermerometric measurements of objects resolved by Rastegaev et al. (2007)} \label{tab2:Rastegaev_n}
\begin{tabular}{l | l | c | c | c | c | c | c | c | c}
\hline
\textbf{Name} & \textbf{Other} & \textbf{Epoch} &$\boldsymbol{\Theta (\degr )}$ & $\sigma_{\Theta}$ & $\boldsymbol{\rho\ ('')}$ & $\sigma_{\rho}$ & $\boldsymbol{\bigtriangleup m}$ & $\sigma_{\boldsymbol{\bigtriangleup m}}$ & $\lambda / \bigtriangleup \lambda$ \\
\textbf{of the object}  & \textbf{designation} &    &    &     &     &    &    &     &  \\
\hline
G102-20             & HIP~26676         & 2006.94164 &  308.0 & 2.8         & 0.119 & 0.006  & 3.2  & 0.4      & 550/20   \\
G191-55             & BD+58$\degr $~876 & 2006.94475 &  125.1 & 0.3         & 0.806 & 0.007  & 2.00 & 0.11     & 800/100  \\
BD+19$\degr $~1185A & HIP~28671         & 2006.94711 &  183.6 & 0.7         & 0.114 & 0.002  & 1.77 & 0.04     & 550/20   \\
G89-14              & HIP~35756         & 2006.94455 &   0.8  & 0.4         & 0.979 & 0.009  & 4.1  & 0.4      & 800/100  \\
G87-45              & NLTT~18038        & 2006.94723 &  271.3 & 0.5         & 0.282 & 0.004  & 2.01 & 0.04     & 550/20   \\
G87-45              & NLTT~18038        & 2006.94724 &  270.7 & 0.4         & 0.282 & 0.003  & 1.76 & 0.04     & 800/100  \\
G87-47              & HIP~36936         & 2006.94725 &  54.0$^{\ast}$ & 2.1 & 0.077 & 0.003  & 1.7  & 0.3      & 800/100  \\
G111-38AB           & HIP~38195         & 2006.94751 &  7.9   & 0.7         & 0.084 & 0.002  & 0.78 & 0.03     & 550/20   \\
G111-38AB           & HIP~38195         & 2006.94749 &  7.8   & 1.3         & 0.084 & 0.002  & 0.75 & 0.03     & 800/100  \\
G111-38AC           & HIP~38195         & 2006.94751 &  200.0 & 0.3         & 2.111 & 0.018  & 1.34 & 0.04     & 550/20   \\
G111-38AC           & HIP~38195         & 2006.94749 &  200.0 & 0.3         & 2.112 & 0.018  & 1.10 & 0.04     & 800/100  \\
G111-38BC           & HIP~38195         & 2006.94751 &  199.5 & 0.3         & 2.193 & 0.019  & 0.57 & 0.05     & 550/20   \\
G111-38BC           & HIP~38195         & 2006.94749 &  199.5 & 0.3         & 2.194 & 0.019  & 0.36 & 0.05     & 800/100  \\
G114-25             & HIP~44111         & 2006.94742 &  323.7 & 0.5         & 0.773 & 0.008  & 3.9  & 0.2      & 800/100  \\
\hline
\multicolumn{7}{l}{$^{\ast}$---the position of the secondary
component is known with $\pm180\degr$ ambiguity.}
\end{tabular}
\end{center}
\end{table*}

\begin{table*}
\begin{center}
\caption{Unresolved stars}\label{unresolved:Rastegaev_n}
\begin{tabular}{l|c|c||l|c|c}
\hline
 \textbf{Name} & \textbf{Filter} {\bf
($\lambda/\Delta\lambda,$nm)} & {\textbf{Epoch}}
 & \textbf{Name} & \textbf{Filter} {\bf
($\lambda/\Delta\lambda,$nm)} & {\textbf{Epoch}} \\ \hline
G265-1       & 550/20; 800/100 & 2007.4952 &   G20-15       & 550/20; 800/100 & 2007.4871 \\
G130-65      & 800/100 & 2007.5147 &           G182-31      & 550/20 & 2007.2493 \\
G31-55       & 600/40; 800/100 & 2007.7253 &   G183-9       & 550/20 & 2007.2494 \\
HD 3567      & 600/40; 800/100 & 2007.7254 &   G183-9       & 600/40 & 2007.5090 \\
G242-65      & 600/40 & 2007.4953 &            G183-11      & 550/20 & 2007.2493 \\
G242-71      & 600/40 & 2007.4952 &            G182-32      & 800/100 & 2007.2437 \\
G271-162     & 800/100 & 2007.7391 &           G183-16      & 550/20 & 2007.2493 \\
BD-1$\degr $ 306 & 550/20; 800/100 & 2007.7391 & G20-24     & 550/20; 800/100 & 2007.4872 \\
G75-31       & 800/100 & 2007.7312 &           G140-44      & 550/20; 800/100 & 2007.4873 \\
G4-36        & 800/100 & 2007.7312 &           G140-46      & 550/20; 800/100 & 2007.4872 \\
G4-37        & 800/100 & 2007.7312 &           G206-34      & 800/100 & 2007.2493 \\
G75-56       & 800/100 & 2007.7366 &           G21-19       & 550/20; 800/100 & 2007.4872 \\
G95-11       & 800/100 & 2007.7367 &           G125-5       & 550/20 & 2007.2493 \\
G89-14       & 550/20 & 2007.2404 &            G92-6        & 600/40; 800/100 & 2007.4901 \\
G13-9        & 550/20 & 2007.2406 &            BD+26$\degr $ 3578   & 550/20; 800/100 & 2007.4901 \\
G11-44       & 550/20 & 2007.2406 &            HD 188510    & 550/20; 800/100 & 2007.4901 \\
G123-9       & 800/100 & 2007.2464 &           G186-26      & 600/40; 800/100 & 2007.4929 \\
G12-21       & 550/20 & 2007.2406 &            HD 194598    & 600/40 & 2007.4954 \\
G13-35       & 550/20 & 2007.2406 &            G262-14      & 600/40; 800/100 & 2007.4953 \\
G13-38       & 550/20 & 2007.2406 &            G24-17       & 600/40 & 2007.4955 \\
G199-20      & 800/100 & 2007.2465 &           G24-17       & 800/100 & 2007.5065 \\
G59-27       & 800/100 & 2007.2464 &           G24-25       & 800/100 & 2007.5064 \\
G60-46       & 550/20 & 2007.2407 &            G210-33      & 800/100 & 2007.5117 \\
G60-48       & 550/20 & 2007.2407 &            G212-7       & 550/20; 800/100 & 2007.5117 \\
G14-33       & 800/100 & 2007.2408 &           HD 201891    & 550/20; 800/100 & 2007.4955 \\
G177-23      & 550/20 & 2007.2465 &            G25-24       & 800/100 & 2007.5065 \\
G255-32      & 800/100 & 2007.2466 &           G187-40      & 800/100 & 2007.5118 \\
G62-52       & 550/20 & 2007.2435 &            G26-1        & 600/40; 800/100 & 2007.4901 \\
G64-12       & 800/100 & 2007.2436 &           G26-1        & 800/100 & 2007.5065 \\
G150-40      & 800/100 & 2007.2464 &           G126-10      & 800/100 & 2007.5093 \\
G165-39      & 550/20 & 2007.2464 &            G93-27       & 800/100 & 2007.5065 \\
G65-22       & 800/100 & 2007.2463 &           G231-52      & 600/40; 800/100 & 2007.4953 \\
G64-37       & 800/100 & 2007.2409 &           G188-22      & 800/100 & 2007.5118 \\
G239-12      & 800/100 & 2007.2466 &           G126-36      & 800/100 & 2007.5066 \\
G178-27      & 550/20 & 2007.2464 &            G188-30      & 800/100 & 2007.5118 \\
G201-5       & 800/100 & 2007.2435 &           G232-40      & 600/40 & 2007.4953 \\
G66-30       & 550/20 & 2007.2410 &            G214-5       & 800/100 & 2007.5118 \\
G166-54      & 800/100 & 2007.2409 &           G27-8        & 800/100 & 2007.5066 \\
G66-51       & 550/20 & 2007.2410 &            G126-52      & 600/40  & 2007.5092 \\
G179-22      & 550/20 & 2007.2465 &            G126-62      & 600/40  & 2007.5092 \\
G201-44      & 550/20 & 2007.2435 &            LFT 1697     & 800/100 & 2007.5066 \\
G15-24       & 800/100 & 2007.2466 &           G18-39       & 800/100 & 2007.5093 \\
G168-26      & 800/100 & 2007.2410 &           G156-7       & 800/100 & 2007.5093 \\
G180-24      & 550/20 & 2007.2434 &            G18-54       & 600/40  & 2007.5093 \\
G202-35      & 800/100 & 2007.2435 &           G27-33       & 800/100 & 2007.5093 \\
G180-58      & 800/100 & 2007.2434 &           G233-26      & 600/40  & 2007.4953 \\
G153-64      & 800/100 & 2007.2438 &           G128-11      & 600/40  & 2007.5094 \\
G17-25       & 550/20; 800/100 & 2007.2438 &   G128-11      & 800/100 & 2007.5119 \\
G202-65      & 800/100 & 2007.2435 &           G242-14      & 600/40  & 2007.4952 \\
G180-66      & 800/100 & 2007.2435 &           G68-3        & 550/20; 800/100 & 2007.5119 \\
G169-28      & 800/100 & 2007.2412 &           G190-15      & 550/20; 800/100 & 2007.5119 \\
G139-8       & 800/100 & 2007.2411 &           G29-25       & 800/100 & 2007.5121 \\
G19-25       & 550/20 & 2007.2494 &            G29-71       & 800/100 & 2007.5121 \\
G139-49      & 550/20 & 2007.2494 &            G20-8        & 550/20 & 2007.2494 \\
\hline
\end{tabular}
\end{center}
\end{table*}

\onecolumn
\newpage
\begin{table*}[bh!]
\begin{center}
\caption{Supplementary data on resolved stars}\label{additional:Rastegaev_n}
\begin{tabular}{l|c|c|c|c}
\hline
\textbf{Name} &  \textbf{Coordinates}       & ${\bf m_V} $& $ {\bf[m/H]^*}$ & \textbf{Total number} \\
\textbf{of the system/subsystem} &   \textbf{(2000.0)} &      &      & \textbf{of components} \\
\hline
G76-21       & $02^h41^m13\fs6$ $+09\degr 46' 12''$ & $10.17$ & $-2.28$   & 2 \\
G89-14       & $07^h22^m31\fs5$ $+08\degr 49' 13''$ & $10.40$ & $-1.90$   & 4 \\
G59-1        & $12^h08^m54\fs7$ $+21\degr 47' 19''$ & $9.49$  & $-1.14$   & 3 \\
G63-46       & $13^h39^m59\fs6$ $+12\degr 35' 22''$ & $9.37$  & $-1.03$   & 2 \\
G135-16      & $14^h04^m01\fs6$ $+22\degr 31' 30''$ & $10.16$ & $-1.04$   & 2 \\
G168-42      & $16^h19^m51\fs7$ $+22\degr 38' 20''$ & $11.51$ & $-1.42$   & 2 \\
G141-47      & $18^h53^m16\fs5$ $+10\degr 37' 26''$ & $10.5$  & $-1.34$   & 2 \\
G142-44      & $19^h38^m53\fs2$ $+16\degr 25' 34''$ & $11.16$ & $-1.17$   & 2 \\
G190-10      & $23^h07^m59\fs8$ $+41\degr 51' 20''$ & $11.22$ & $-1.92$   & 3 \\
G28-43       & $23^h09^m32\fs9$ $+00\degr 42' 40''$ & $9.96$  & $-1.80$   & 2 \\
G217-8       & $23^h26^m32\fs8$ $+60\degr 37' 43''$ & $10.47$ & $-2.24$   & 2 \\
G130-7       & $23^h45^m00\fs1$ $+30\degr 20' 10''$ & $11.72$ & $-1.62$   & 2 \\
\hline
\multicolumn{5}{l}{$^{\ast}$---metallicities are adopted from the
CLLA catalog (\cite{clla:Rastegaev_n}).}
\end{tabular}
\end{center}
\end{table*}

\newpage
\onecolumn
\begin{center}
\begin{longtable}{l|c|c|c|c|c}
\caption{WDS components for the stars of the sample}
\label{wds:Rastegaev_n}\\
\hline
\textbf{Name} & \textbf{WDS companion} & $\boldsymbol{\rho\ ('')}$ & $\boldsymbol{\bigtriangleup m}$& \textbf{Status} & \textbf{References} \\
\hline \hline
\endfirsthead
\caption{WDS components of the stars of the sample (Contd.)} \\
\hline
\textbf{Name} &  \textbf{WDS companion} & $\boldsymbol{\rho\ ('')}$& $\boldsymbol{\bigtriangleup m}$& \textbf{Status} & \textbf{Reference} \\
\hline \hline
\endhead
\hline
\endfoot
G242-65      & 00437+7211OSO  10AB   &   &   & $-$ & (\cite{osorio:Rastegaev_n}) \\
             & 00437+7211OSO  10AC   &   &   & $-$ &
(\cite{osorio:Rastegaev_n}) \\
G59-1        & 12089+2147HDS1714Aa   & $0.3$  & $2.25$ & $+$ & (\cite{esa:Rastegaev_n}), $\star$ \\
             & 12089+2147LDS 930AB   & $15.7$ & $5.51$ & $+$ &
(\cite{allen:Rastegaev_n}) \\
G62-52       & 13360+0112OSO  54     &   &   & $-$ & (\cite{osorio:Rastegaev_n}) \\
G63-46       & 13400+1235HDS1917     & $0.2$  & $0.68$  &  $+$ & (\cite{esa:Rastegaev_n}), $\star$  \\
G239-12      & 14189+7314OSO  55AB   &   &   & $-$ & (\cite{osorio:Rastegaev_n}) \\
             & 14189+7314OSO  55AC   &   &   & $-$ &
(\cite{osorio:Rastegaev_n}) \\
G179-22      & 15144+3301OSO  62     &   &   & $-$ & (\cite{osorio:Rastegaev_n}) \\
G15-24       & 15307+0824OSO  64     &   &   & $-$ & (\cite{osorio:Rastegaev_n}) \\
G180-24      & 16032+4215OSO  67     &   &   & $-$ & (\cite{osorio:Rastegaev_n}) \\
G168-42      & 16199+2238OSO  68AB   &   &   & $-$ & (\cite{osorio:Rastegaev_n}) \\
             & 16199+2238OSO  68AC   &   &   & $-$ &
(\cite{osorio:Rastegaev_n}) \\
G180-58      & 16283+4441OSO  71     &   &   & $-$ & (\cite{osorio:Rastegaev_n}) \\
G153-64      & 16325-0834OSO  72     &   &   & $-$ & (\cite{osorio:Rastegaev_n}) \\
G17-25       & 16348-0412GIC 144AB   &$1170.7$&$4.25$& $+$ & (\cite{osorio:Rastegaev_n}) \\
             & 16348-0412LMP  14AC   &   &   & $-$ &
(\cite{osorio:Rastegaev_n}) \\
             & 16348-0412LMP  14BD   &   &   & $-$ &
(\cite{osorio:Rastegaev_n}) \\
             & 16348-0412LMP  14BE   &   &   & $-$ &
(\cite{osorio:Rastegaev_n}) \\
G169-28      & 16502+2219OSO  74AB   &   &   & $-$ & (\cite{osorio:Rastegaev_n}) \\
             & 16502+2219OSO  74AC   &   &   & $-$ &
(\cite{osorio:Rastegaev_n}) \\
G19-25       & 17260-0245OSO  78AB   &   &   & $-$ & (\cite{osorio:Rastegaev_n}) \\
             & 17260-0245OSO  78AC   &   &   &  ?  & \\
G20-8        & 17398+0225OSO  83AB   &   &   & $-$ & (\cite{osorio:Rastegaev_n}) \\
             & 17398+0225OSO  83AC   &   &   & $-$ &
(\cite{osorio:Rastegaev_n}) \\
G20-15       & 17475-0847OSO  84AB   &   &   & $-$ &  \\
             & 17475-0847OSO  84AC   &   &   & $-$ &  \\
             & 17475-0847OSO  84AD   &   &   & $-$ &  \\
G182-31      & 17523+3624OSO  85     &   &   & $-$ & (\cite{osorio:Rastegaev_n}) \\
G183-9       & 17530+1521OSO  86AB   &   &   & $-$ & (\cite{osorio:Rastegaev_n}) \\
             & 17530+1521OSO  86AC   &   &   & $-$ &
(\cite{osorio:Rastegaev_n}) \\
G183-11      & 17547+2016OSO  88     &   &   & $-$ & (\cite{osorio:Rastegaev_n}) \\
G182-32      & 17551+3745OSO  89AB   &   &   & $-$ & (\cite{osorio:Rastegaev_n}) \\
             & 17551+3745OSO  89AC   &   &   & $-$ &
(\cite{osorio:Rastegaev_n}) \\
             & 17551+3745OSO  89AD   &   &   & $-$ &
(\cite{osorio:Rastegaev_n}) \\
G20-24       & 18079+0153OSO  93AB   &   &   &  ?  &  \\
             & 18079+0153OSO  93AC   &   &   & $-$ &
(\cite{osorio:Rastegaev_n}) \\
             & 18079+0153OSO  93AD   &   &   & $-$ &
(\cite{osorio:Rastegaev_n}) \\
             & 18079+0153OSO  93AE   &   &   & $-$ &
(\cite{osorio:Rastegaev_n}) \\
             & 18079+0153OSO  93AF   &   &   & $-$ &  \\
G140-44      & 18115+1455OSO  94     &   &   & $-$ & (\cite{osorio:Rastegaev_n}) \\
G140-46      & 18124+0524OSO  95     &   &   & $-$ & (\cite{osorio:Rastegaev_n}) \\
G206-34      & 18353+2842OSO 101AB   &   &   & $-$ & (\cite{osorio:Rastegaev_n}) \\
             & 18353+2842OSO 101AC   &   &   & $-$ &
(\cite{osorio:Rastegaev_n}) \\
             & 18353+2842OSO 101AD   &   &   & $-$ &
(\cite{osorio:Rastegaev_n}) \\
             & 18353+2842OSO 101AE   &   &   & $-$ &
(\cite{osorio:Rastegaev_n}) \\
             & 18353+2842OSO 101AF   &   &   & $-$ &
(\cite{osorio:Rastegaev_n}) \\
G92-6        & 19297+0102OSO 109AB   &   &   & $-$ & (\cite{osorio:Rastegaev_n}) \\
             & 19297+0102OSO 109AC   &   &   & $-$ &
(\cite{osorio:Rastegaev_n}) \\
             & 19297+0102OSO 109AD   &   &   &  ?  &  \\
             & 19297+0102OSO 109AE   &   &   &  ?  &  \\
             & 19297+0102OSO 109AF   &   &   & $-$ &
(\cite{osorio:Rastegaev_n}) \\
             & 19297+0102OSO 109AG   &   &   &  ?  &  \\
             & 19297+0102OSO 109AH   &   &   & $-?$&  \\
             & 19297+0102OSO 109AI   &   &   &  ?  &  \\
             & 19297+0102OSO 109AJ   &   &   & $-$ &
(\cite{osorio:Rastegaev_n}) \\
             & 19297+0102OSO 109AK   &   &   &  ?  &  \\
G142-44      & 19389+1626OSO 110AB   &   &   & $-$ &  \\
             & 19389+1626OSO 110AC   &   &   & $-$ &
(\cite{osorio:Rastegaev_n}) \\
             & 19389+1626OSO 110AD   &   &   & $-$ &
(\cite{osorio:Rastegaev_n}) \\
             & 19389+1626OSO 110AE   &   &   & $-$ &
(\cite{osorio:Rastegaev_n}) \\
             & 19389+1626OSO 110AF   &   &   & $-$ &
(\cite{osorio:Rastegaev_n}) \\
             & 19389+1626OSO 110AG   &   &   & $-?$&  \\
             & 19389+1626OSO 110AH   &   &   & $-$ &
(\cite{osorio:Rastegaev_n}) \\
G186-26      & 20248+2503OSO 125AB   &   &   &  ?  & \\
             & 20248+2503OSO 125AC   &   &   & $-?$& \\
G210-33      & 20454+4023OSO 133AB   &   &   & $-$ & (\cite{osorio:Rastegaev_n}) \\
             & 20454+4023OSO 133AC   &   &   & $-$ &
(\cite{osorio:Rastegaev_n}) \\
             & 20454+4023OSO 133AD   &   &   & $-$ &
(\cite{osorio:Rastegaev_n}) \\
             & 20454+4023OSO 133AE   &   &   & $-$ &
(\cite{osorio:Rastegaev_n}) \\
G212-7       & 20553+4218OSO 137AB   &   &   & $-$ & (\cite{osorio:Rastegaev_n}) \\
             & 20553+4218OSO 137AC   &   &   & $-$ &
(\cite{osorio:Rastegaev_n}) \\
             & 20553+4218OSO 137AD   &   &   & $-$ &
(\cite{osorio:Rastegaev_n}) \\
             & 20553+4218OSO 137AE   &   &   &  ?  &  \\
             & 20553+4218OSO 137AF   &   &   & $-$ &
(\cite{osorio:Rastegaev_n}) \\
             & 20553+4218OSO 137AG   &   &   & $-$ &
(\cite{osorio:Rastegaev_n}) \\
G187-40      & 21220+2727OSO 145     &   &   & $-$ & (\cite{osorio:Rastegaev_n}) \\
G93-27       & 21399+0623OSO 151AB   &$3.3$& $2.66$  & $+$ & (\cite{osorio:Rastegaev_n}) \\
             & 21399+0623OSO 151AC   &   &   & $-$ &
(\cite{osorio:Rastegaev_n}) \\
G231-52      & 21393+6017OSO 150     &   &   & $-$ & (\cite{osorio:Rastegaev_n}) \\
G188-22      & 21440+2723OSO 155     &$5.0$&$6.93$& $+$ & (\cite{osorio:Rastegaev_n}) \\
G188-30      & 21553+3239OSO 162AB   &   &   & $-$ & (\cite{osorio:Rastegaev_n}) \\
             & 21553+3239OSO 162AC   &   &   & $-$ &
(\cite{osorio:Rastegaev_n}) \\
G232-40      & 21554+5608OSO 163AB   &   &   & $-$ & (\cite{osorio:Rastegaev_n}) \\
             & 21554+5608OSO 163AC   &   &   & $-$ &
(\cite{osorio:Rastegaev_n}) \\
G214-5       & 21592+4102OSO 164AB   &   &   & $-$ & (\cite{osorio:Rastegaev_n}) \\
             & 21592+4102OSO 164AC   &   &   & $-$ &
(\cite{osorio:Rastegaev_n}) \\
             & 21592+4102OSO 164AD   &   &   & $-$ &
(\cite{osorio:Rastegaev_n}) \\
G27-8        & 22032-0113LDS4938AB   &   &   & $-?$&  \\
             & 22032-0113OSO 166AC   &   &   &  ?  &  \\
G126-62      & 22115+1806CHR 119Aa,Ab &$0.2$ &   & $+?$&  \\
             & 22115+1806OSO 171Aa,B &   &   & $-$ &
(\cite{osorio:Rastegaev_n}) \\
LFT 1697     & 22144-0845OSO 174     &   &   & $-$ & (\cite{osorio:Rastegaev_n}) \\
G18-39       & 22186+0827OSO 175     &   &   & $-$ & (\cite{osorio:Rastegaev_n}) \\
G27-33       & 22328-0557OSO 181     &   &   & $-$ & (\cite{osorio:Rastegaev_n}) \\
G233-26      & 22399+6143OSO 184AB   &   &   & $-$ & (\cite{osorio:Rastegaev_n}) \\
             & 22399+6143OSO 184AC   &   &   & $-$ &
(\cite{osorio:Rastegaev_n}) \\
G190-10      & 23080+4151OSO 189     &   &   & $-$ & (\cite{osorio:Rastegaev_n}) \\
G217-8       & 23265+6038OSO 196AB   &   &   & $-$ & (\cite{osorio:Rastegaev_n}) \\
             & 23265+6038OSO 196AC   &   &   & $-$ &
(\cite{osorio:Rastegaev_n}) \\
             & 23265+6038OSO 196AD   &   &   & $-$ &
(\cite{osorio:Rastegaev_n}) \\
             & 23265+6038OSO 196AE   &   &   & $-$ &
(\cite{osorio:Rastegaev_n}) \\
G130-7       & 23450+3020OSO 204     &   &   & $-$ & (\cite{osorio:Rastegaev_n}) \\
G29-71       & 23500+0843OSO 207     &   &   & $-$ & (\cite{osorio:Rastegaev_n}) \\
\end{longtable}
\end{center}

\end{document}